\documentclass{iauc}
\usepackage{graphicx}
\pubyear{2005}
\volume{199}
\pagerange{1--}
\jname{Probing Galaxies through Quasar Absorption Lines}
\editors{P. R. Williams, C. Shu, and B. M\'{e}nard, eds.}

\title[The QSO Absorber - Galaxy Connection] 
{The Connections between QSO Absorption Systems and Galaxies: Low-Redshift Observations}

\author[Tripp and Bowen]   
{Todd M. Tripp$^1$%
\and David V. Bowen$^2$}

\affiliation{$^1$Department of Astronomy, University of Massachusetts,
Amherst, MA 01003, USA \break 
email: tripp@astro.umass.edu\\[\affilskip]
$^2$Princeton University Observatory, Princeton, NJ 08544, USA \break 
email: dvb@astro.princeton.edu}

\begin{document}
\newcommand{\apg}       {^{>}_{\sim}}
\newcommand{\h}         {$h^{-1}_{70}\,$~kpc}

\maketitle

\begin{abstract}
Quasar absorption lines have long been recognized to be a sensitive
probe of the abundances, physical conditions, and kinematics of gas in
a wide variety of environments including low-density intergalactic
regions that probably cannot be studied by any other means. While some
pre-{\it Hubble Space Telescope (HST)} observations indicated that
Mg~II absorption lines arise in gaseous galactic halos with a large
covering factor, many early QSO absorber studies were hampered by a
lack of information about the context of the absorbers and their
connections with galaxies. By providing access to crucial ultraviolet
resonance lines at low redshifts, deployment of {\it HST} and the {\it
Far Ultraviolet Spectroscopic Explorer} enabled detailed studies of
the relationships between QSO absorbers and galaxies. The advent of
large surveys such as the {\it Sloan Digital Sky Survey (SDSS)} has
also advanced the topic by greatly improving the size of absorber and
galaxy samples. This paper briefly reviews some observational results
on absorber-galaxy connections that have been obtained in the {\it
HST/SDSS} era, including Mg~II absorbers, the low$-z$ Ly$\alpha$
forest, Lyman limit and damped Ly$\alpha$ absorbers, and O~VI systems.

\keywords{quasars: absorption lines; galaxies: evolution, formation, ISM, halos}

\end{abstract}

\firstsection 
\section{Introduction}

\subsection{Some Advantages of QSO Absorption Lines}

\begin{figure}
\includegraphics[height=7in]{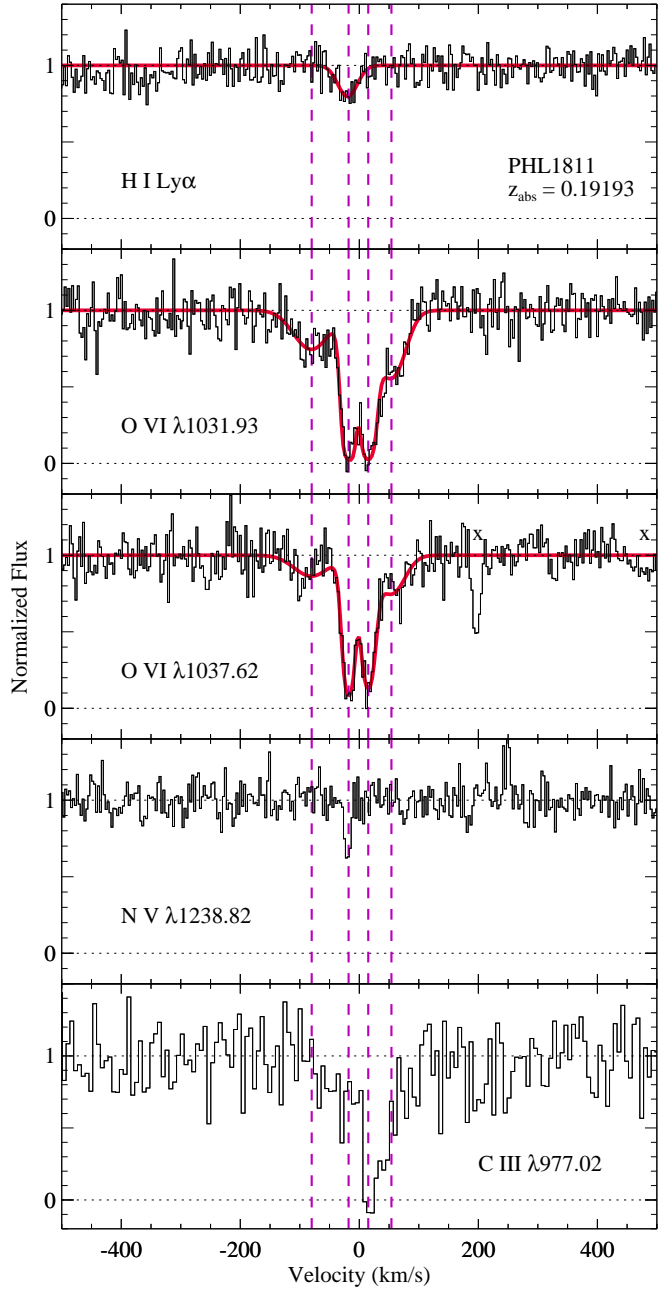}
\caption{Complex, multicomponent O~VI absorption system detected in
the UV spectrum of PHL1811 obtained by Jenkins et al. (2005) with the
STIS E140M echelle spectrograph. The panels show the
continuum-normalized profiles of several species of interest plotted
vs velocity in the absorber frame ($v$ = 0 at $z_{\rm abs}$ =
0.19193). Multiple components are unambiguously detected in this
absorber, and moreover, the physical conditions and/or metallicity of
the gas change dramatically from one component to the next (compare
the relative strengths of H~I, C~III, Si~III, N~V, and O~VI). The high
spectral resolution provided by STIS is crucial for successful
analysis of complex absorption systems such as this
one.\label{phl1811example}}
\end{figure}

Absorption-line spectroscopy is a powerful tool for the study of gas
in the universe. A wide array of elements can be detected in
absorption; in the Milky Way ISM, absorption lines ranging from common
elements (e.g., H, C, N, O) to exotic heavy metals (e.g., arsenic,
selenium, tin, and lead, see, e.g., Cardelli et al. 1991,1993; Hobbs
et al. 1993) have been detected. With adequate signal-to-noise, these
exotic species can in principle be observed in higher column density
QSO absorbers as well (Prochaska, Howk, \& Wolfe 2003). Abundance
patterns in QSO absorption systems provide valuable information on the
nucleosynthetic origins and dust content of the gas and the buildup of
metals vs. time. QSO absorption lines also probe the physical
conditions of interstellar and intergalactic material: line widths,
excited fine-structure lines, and column density ratios yield good
constraints on the temperature, density, pressure, and ionization
mechanism as well as the radiation field to which the gas is exposed.

Possibly the greatest advantage of absorption spectroscopy, however,
is its tremendous sensitivity.  QSO absorption lines can be used to
detect low-density gas that is orders of magnitude below the detection
threshold of most other techniques. As an example,
Figure~\ref{phl1811example} shows absorption lines detected at $z_{\rm
abs}$ = 0.19193 in a UV echelle spectrum of the bright QSO PHL1811
(Jenkins et al. 2003) obtained in a modest {\it HST} program with the
Space Telescope Imaging Spectrograph (STIS). The H~I Ly$\alpha$
absorption line shown in the top panel of Figure~\ref{phl1811example}
has log $N$(H~I) = 12.8 and is detected at the 4$\sigma$ level. For
comparison, 21 cm emission can be used to study H~I-bearing gas, and
with considerable effort, gas with log $N$(H~I) $>$ 18.0 can be
detected in 21 cm emission.  For the study of H~I, high-resolution
spectra are typically five or six orders of magnitude more sensitive
than 21 cm emission observations.  Moreover, absorption lines can be
measured comparably well from $z = 0$ out to $z >$ 4 while 21 cm
emission can only be detected in the nearby universe with current
facilities.  And, absorption is not limited to low-density, weak
absorption lines; high column densities can be measured with great
precision as well because these cases show broad damping wings spread
over many pixels.  Intrinsically weak lines (e.g., spin-changing
transitions such as the O~I 1355.6 \AA\ line) can also be used to
measure abundances in high column density and/or high metallicity
systems in which the usual resonance lines are strongly saturated but
not strong enough to show damping wings.

However, QSO absorption systems can be complex, and in order to take
full advantage of the technique, high spectral resolution and broad
wavelength coverage are required.  Figure~\ref{phl1811example} again
provides good examples of absorber complexity and the benefits of high
resolution with broad coverage.  Comparing the various panels in
Figure~\ref{phl1811example}, it is readily apparent that there is more
to life than H~I -- Ly$\alpha$ only reveals the tip of the iceberg in
this system. Four components are evident in the O~VI doublet at $v =
-80, -18, 15,$ and +54 km s$^{-1}$. Ly$\alpha$ is only detected at
$-18$ km s$^{-1}$ along with N~V, weak C~III, and O~VI. This is also
the only component to show N~V.  At first glance, C~III appears to be
strongest at yet a different velocity.  However, because the STIS
spectrum covers a substantial wavelength range, we realize that a
strong Ly$\alpha$ system is present at a different redshift, one that
shifts its Ly$\beta$ line into a blend with the C~III line shown in
Figure~\ref{phl1811example}. Thanks to the high spectral resolution of
the data, we can decompose this C~III/Ly$\beta$ blend and at least
provide useful constraints based on the C~III line.  Indeed, at lower
resolution, these lines in Figure~\ref{phl1811example} could still be
detected (with adequate signal-to-noise), but such data would not
allow the column densities to be correctly assigned to different
components, which could lead to erroneous conclusions. High spectral
resolution provides many other benefits, e.g., better ability to
assess and correct for line saturation and better ability to measure
widths of narrow lines such the N~V line in
Figure~\ref{phl1811example}.

\subsection{Some Disadvantages of QSO Absorption Lines}

However, like all observational techniques, QSO absorption lines have
some drawbacks. For example, faint QSOs are difficult to observe at
high resolution -- especially at lower redshifts -- which limits
sample sizes. The main disadvantage, however, is that the technique
only provides information about material along the pencil beam to the
background QSO. At higher redshifts, it is extremely challenging to
understand how the detailed information provided by quasar absorption
measurements is really connected with the more global context, e.g.,
whether the absorption arises in the IGM or the ISM of a galaxy, what
type of galaxy, etc.  However, at low redshifts, detailed information
about the environment in which the absorption occurs can be obtained,
information such as redshifts and deep images of nearby galaxies, 21
cm or CO emission maps, and the locations of nearby galaxy
groups/clusters, voids, and large-scale structures. With such
knowledge, we can hope to understand the real nature of QSO absorption
lines and their implications for galaxy evolution and cosmology.  

The purpose of this paper is to present some examples of how studies
of the connections between low-redshift absorbers and galaxies have so
far advanced our understanding of the topic as well as goals for
future work.  This brief (and incomplete) review will comment on
studies of low$-z$ Mg~II absorbers, Ly$\alpha$ clouds, systems with
higher H~I column densities, and O~VI absorbers.

\section{Mg~II Absorbers\label{secmg2}}

Historically, the study of QSO absorber-galaxy relationships began
with Mg~II absorption systems. Magnesium is not particularly abundant
and moreover is prone to confusion from dust depletion and ionization
effects, so in hindsight, Mg~II seems like a strange species to use
for probing the galaxy-absorber connection. However, this choice was
strongly driven by practical constraints: before the deployment of
{\it HST}, only Mg~II could be studied from the ground at redshifts
low enough to allow follow-up investigation of nearby galaxies. Among
the dominant ions in H~I gas, Mg~II is the species with the
longest-wavelength resonance transitions, the Mg~II doublet at
$\lambda \lambda$2796.35, 2803.53. Intrinsically more useful species
(e.g., H~I or O~I) only have resonance transtions at $\lambda \ll$
2000 \AA , and ions with resonance lines in the optical (e.g., Na~I or
Ca~II) are even more difficult to analyze than Mg II.  

The first observational test of the Mg~II - galaxy connection was to
image the field around a QSO with a known Mg~II system and then
measure the redshifts of galaxies close to the line of sight to search
for objects at the redshift of the absorption system. Such
observations were pioneered by Bergeron (1986), Bergeron~\etal\
(1987), Cristiani (1987), and Bergeron \& Boiss\'{e} (1991). These
early studies successfully found galaxies at the Mg~II absorption
redshifts and at relatively small impact parameters. The investigation
of Mg~II systems with the largest sample, however, was done by Steidel
and collaborators, who found that all normal galaxies with
luminosities $L^*(B)$ had Mg~II absorbing halos of radius $R^{*}\sim
60$~\h , and that the radius of a halo scales weakly with galaxy
luminosity (Steidel et al.~1994; Steidel 1994).  These studies
appeared to establish a direct link between an individual galaxy and a
Mg~II absorption system. However, the nature of this link remains
unclear. Steidel et al. (2002) have compared the kinematics of five
Mg~II absorbers with the rotation curves of nearby ($\rho \leq 110
h_{70}^{-1}$ kpc) inclined galaxies, and they conclude that most of
these Mg~II systems arise in rotating halo gas associated with the
nearby galaxy. On the other hand, Chuchill, Steidel, \& Vogt (1996)
have compared high-resolution Mg~II absorption recordings to
high-quality {\it HST} observations of nearby galaxies, and they find
no compelling correlations between the absorber properties and the
nearby galaxy properties. Further comments and more recent results on
Mg~II absorber - galaxy relationship studies are found in Churchill,
Kacprzak, \& Steidel (2005). It now appears likely that Mg~II systems
have a variety of origins, and more observations are needed.

More vigorous comparison of observations to theoretical work on gas in
galaxies is also required. Over the last decade, theorists have
produced detailed hydrodynamical simulations which trace the growth of
large-scale structure in the universe and predict the emergence of a
`cosmic web' of filaments connected at nodes where galaxy clusters are
found. The filamentary distribution of galaxies has been abundantly
verified by large galaxy redshift surveys such as 2MASS (e.g., Maller
et al. 2003), SDSS (e.g., Stoughton et al. 2002), and 2dFGRS (e.g.,
Colless et al. 2001), but observations of gas in the web are much more
sparse.  Indeed, these models would seem to require a re-assesment of
how absorbing gas and galaxies might be connected. In the dark matter
halos and filaments, gas and galaxies share the same gravitational
potentials --- and so would be found at similar redshifts --- but the
exact interplay between a galaxy and the intergalactic medium (IGM) is
poorly understood. On the one hand, the $z<1$ IGM may be
metal-enriched through much earlier episodes of star-formation
[e.g. winds from the first galaxies (Aguirre~\etal\ 2001), population
III stars, etc.] so the presence of a galaxy is not a necessary
condition for the existence of metallic (absorbing) gas.  On the other
hand, galaxies probably continue to enrich the IGM from their
formation through to the present (Heckman~\etal\ 2000) via a variety
of processes, including: winds from starburst and blue compact dwarf
galaxies, the more normal expulsion of interstellar gas via galactic
chimneys, tidal stripping through galaxy-galaxy interactions, the
evaporation of dwarf galaxies, or ram-pressure stripping from passage
through an intragroup/intracluster medium.

So what do Mg~II systems represent in this new paradigm of galaxies
embedded in the cosmic web?  There seems little doubt that Galactic
{\it disks} give rise to strong Mg~II lines since disks have a
substantial $N$(H~I) and relatively high metallicity over a large
cross-section of a galaxy.  Our own Milky Way, for example, produces
complex, saturated Mg~II absorption along all sightlines which
penetrate (only half of) the Galactic disk (e.g., Bowen et al. 1995;
Savage et al. 2000). However, several groups have concluded that
rotating disks by themselves are not sufficient to explain the
observed Mg~II kinematics (e.g., Charlton \& Churchill 1998; Steidel
et al. 2002). Some Mg~II absorbers are strong and strikingly
symmetric, and this has led to suggestions that these particular
systems arise in superbubbles/superwinds (Bond et al. 2001; Ellison,
Mallen-Ornelas, \& Sawicky 2003).

\subsection{Do  all galaxies cause Mg~II absorption?}

\begin{figure}
\includegraphics[height=8.5cm,angle=0]{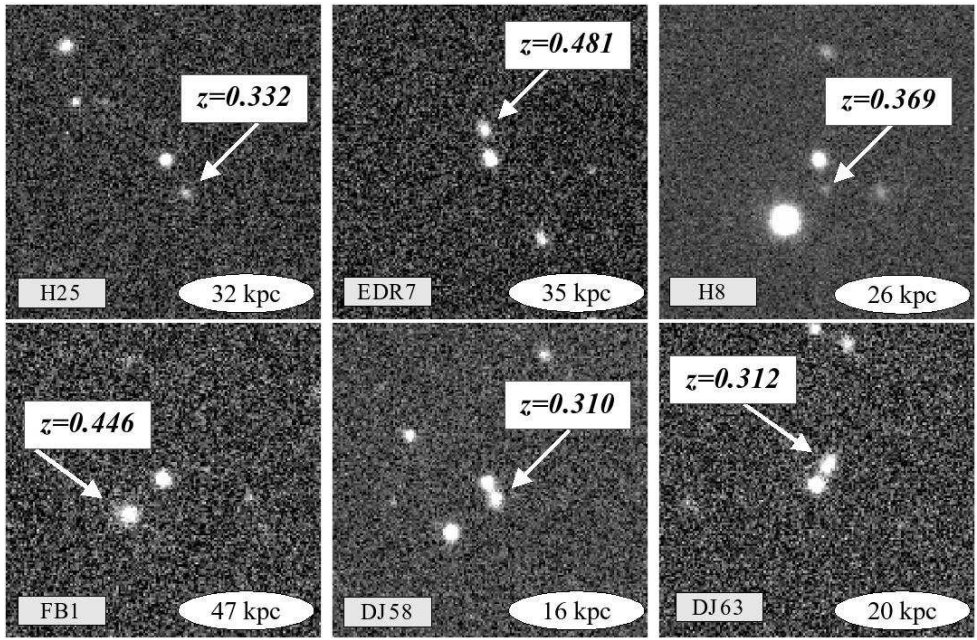}
\caption{Sections from six SDSS $r$-band images showing some of the
  QSO-galaxy pairs studied. The background QSO is the bright stellar-like
  source at the center of each image, and the intervening foreground galaxy is
  identified with an arrow. The measured redshifts are indicated. The
  separation between the galaxy and the QSO sightline is indicated
  at the bottom-right of each panel, in units of \h . Each image is 1 arcmin on a
  side. \label{fig_panel} }
\end{figure}

This is a fundamental first question. Mg~II studies in the late 1980s
and early 1990s gave the impression that if a luminous galaxy is
within some impact parameter of a background QSO sight line, Mg~II
absorption would almost always be detected. This is somewhat
surprising because in the Milky Way, interstellar clouds are small and
have a patchy distribution. Most of these early searches for absorbing
galaxies were conducted toward QSOs with Mg~II systems {\it which were
already known}. Hence, although Mg~II absorbers appear to be
associated with galaxies in some way, we do not know that {\it all}
galaxies (and/or their environments) give rise to Mg~II systems. To
address this question, it is worthwhile to invert the procedure
originally used to study Mg~II-galaxy relationships, i.e., {\it first}
select galaxies close to QSO sightlines (without any knowledge of the
absorption spectrum), {\it then} search for Mg~II lines at the
foreground galaxy redshift.

This inverted experiment is useful for testing ideas about how
galaxies and the IGM might be connected.  If galaxies and QSO
absorbers are only loosely associated, the detection rate and covering
factor of Mg~II around galaxies should be considerably smaller than
the 100\% found from the original studies. And there should be no
correlation with galaxy luminosity.  Clearly, if the gas is merely a
component of the IGM, then it will be an important challenge to
understand the physical state (ionization, kinematics and abundances)
of such lone clouds. As we discuss below, there are many viable
explanations for a low covering factor for Mg~II absorbing gas.
Nevertheless, a covering factor substantially less than 100\% would
suggest that the origin of Mg~II systems is more complicated than
originally thought.  Alternatively, if Mg~II absorbers do have a
near-unity covering factor and adhere to the scaling relationship
between radius and luminosity found in the earlier studies, then
modeling of galaxy evolution and dynamics would have to take into
account the existence of extended and metal-enriched baryonic halos
and explain how galaxies can sustain such halos over a significant
fraction of their lifetime.

Selecting a significant number of $z\:>\:0.2$ galaxies which lie close
on the plane of the sky to QSOs (without forehand knowledge of
absorption in the QSO spectrum) is challenging because the galaxies
need to be found within $\sim 10''$ of a sightline to be within a
proper distance of $\sim$60~\h . The QSO must also be bright enough to
be observed spectroscopically at a resolution of only a few \AA\ for
the follow-up search for Mg~II.  Selection of such QSO-galaxy pairs
has, until recently, required prohibitively large amounts of observing
time. Now, however, the {\it SDSS} has made such an experiment
possible.  The wide area of sky covered, the robust identification of
large numbers of QSOs, and the availibility of multicolor photometry
for selection of foreground galaxies ensures that substantial
numbers of potential QSO-galaxy pairs can be investigated.

\subsection{Results}

\begin{figure}
\includegraphics{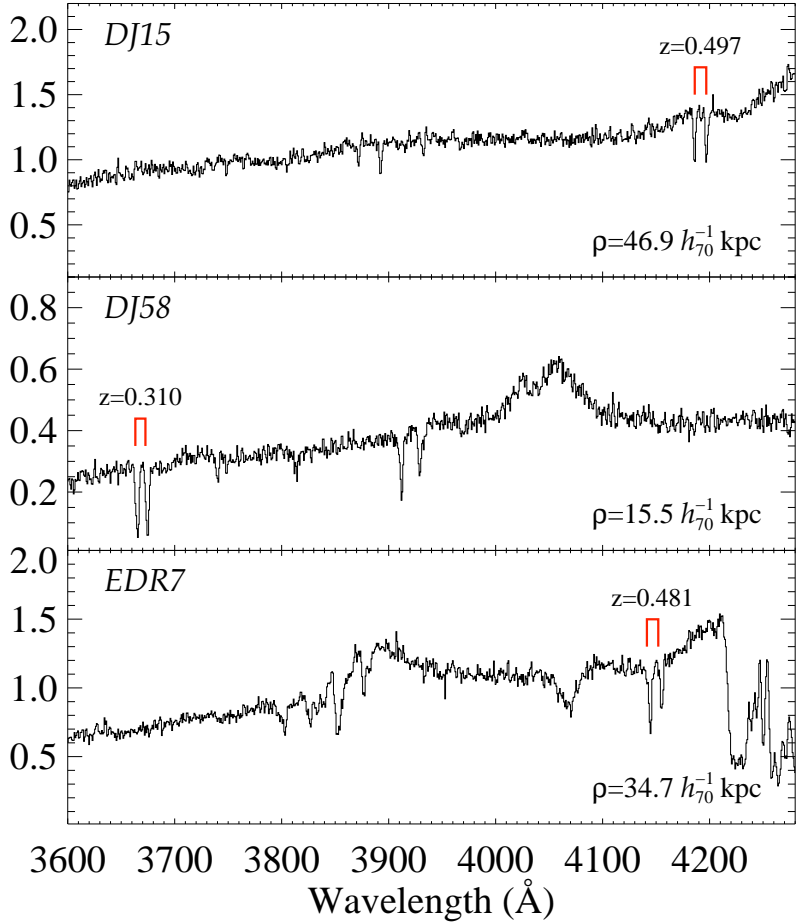}
\caption{Example {\it Multi Mirror Telescope} spectra of QSOs whose
 sightlines pass close to foreground galaxies. This figure shows
 relative flux vs. observed wavelength for three cases where Mg~II
 absorption is detected from the foreground galaxy. The expected
 wavelengths of the Mg~II doublet at the redshifts of the intervening
 galaxies are indicated above each spectrum, and the galaxy-sight line
 separation is listed at the bottom right of each
 panel. \label{fig_spec_yes}}
\end{figure}

\begin{figure}
\includegraphics{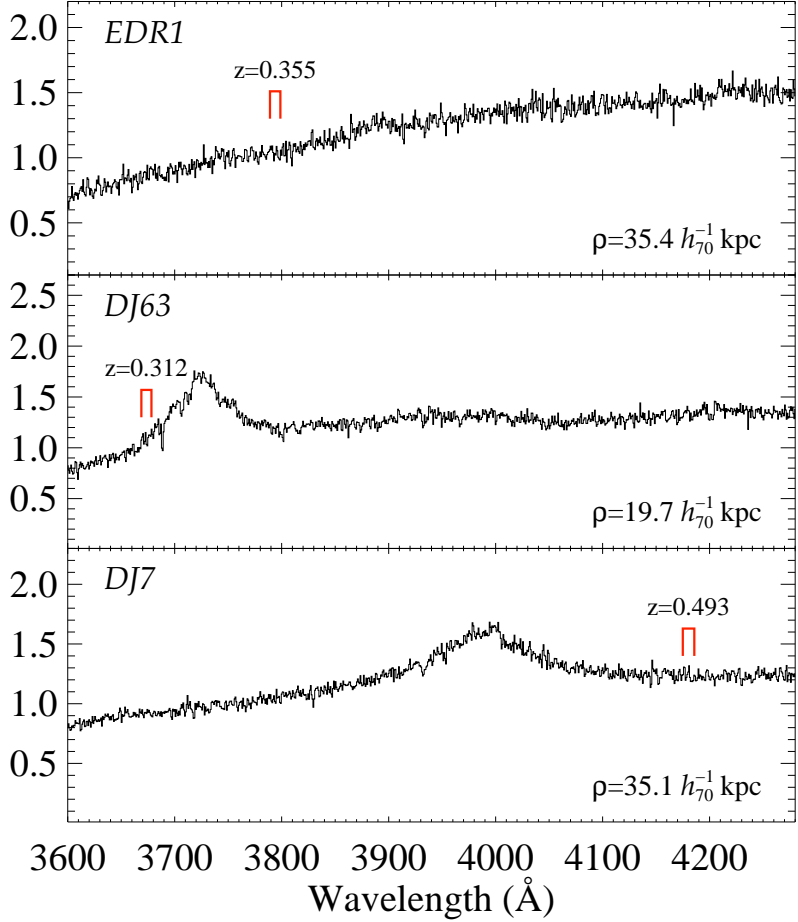}
\caption{Example {\it Multi Mirror Telescope} spectra of QSOs whose
 sightlines pass close to foreground galaxies that {\it do not} show
 Mg~II absorption. As in Figure~\ref{fig_spec_yes}, the redshifts of
 the intervening galaxies are indicated, and their separation from the
 QSO sightline is given at the bottom right of each
 spectrum. \label{fig_spec_no}}
\end{figure}

\begin{figure}
\includegraphics[height=7cm]{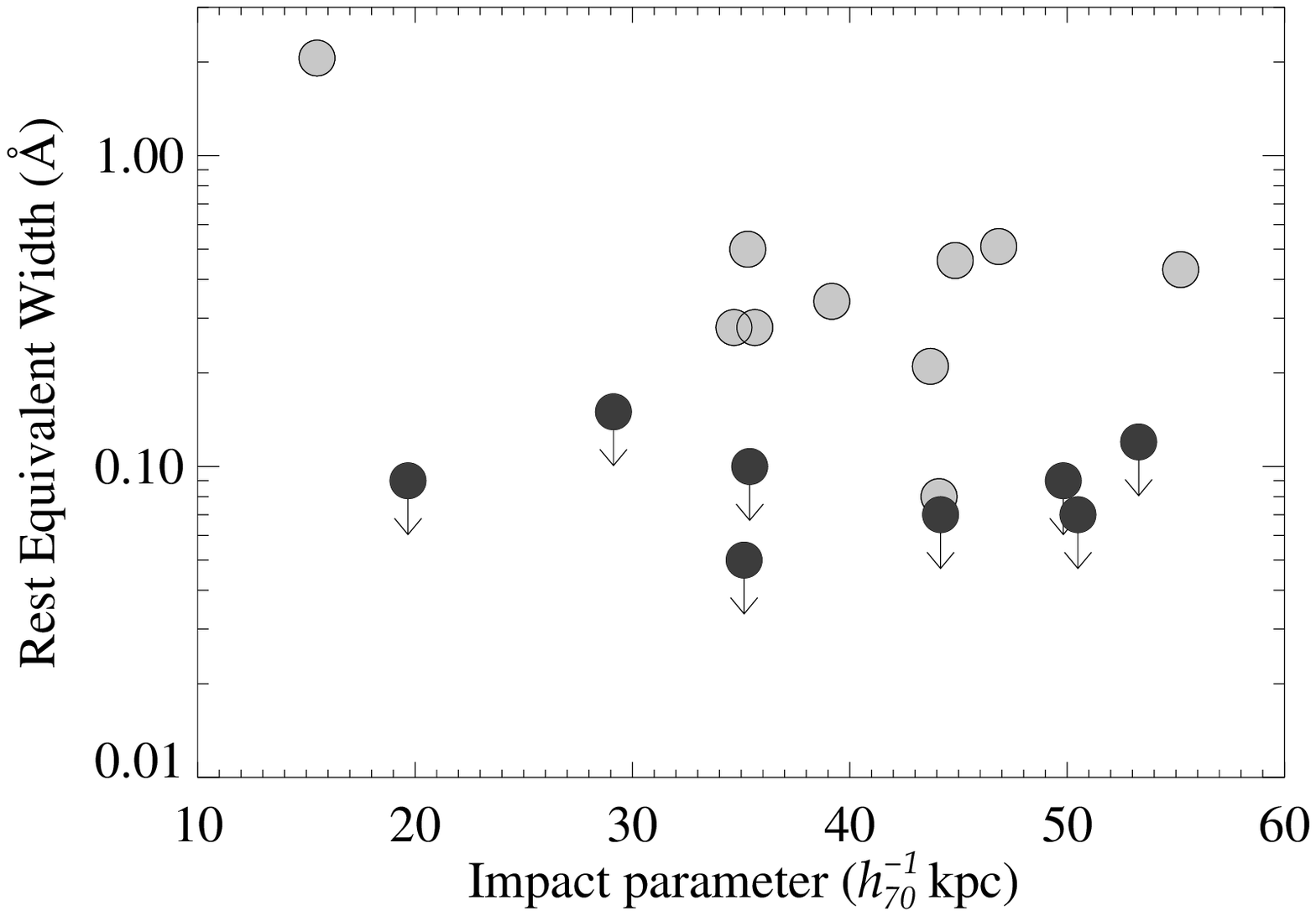}
\caption{The rest equivalent width of the Mg~II~$\lambda 2796$ absorption
line --- or the $2\sigma$ upper  limit --- at the redshift of the intervening
galaxy vs.\ impact parameter. \label{fig_ew}}
\end{figure}

We have been taking advantage of the {\it SDSS} in order to carry out
the inverted Mg~II - galaxy study described above. Briefly, we have
used multicolor {\it SDSS} imaging to first select galaxy-QSO pairs;
photometric redshifts were used to select galaxies with $z > 0.2$ (so
that Mg~II could be observed from the ground). While SDSS does obtain
follow-up spectroscopy, the galaxies and QSOs in the selected pairs
were generally too faint to be covered in the spectroscopic component
of the SDSS. Consequently, we have used the {\it Hobby Ebberly
Telescope} to measure the redshifts of thirty $z=0.31-0.55$ galaxies
within $14-51$~\h\ of $g < 20$ QSOs.  Fig.~\ref{fig_panel} shows
examples of the SDSS $r$-band images (from {\it Data Release 3}) of
six such pairs. After the galaxy redshifts were securely measured, in
December 2004 we completed a substantial fraction of the second phase
of the experiment by using the {\it Multi Mirror Telescope (MMT)} to
observe two thirds of the QSOs at intermediate spectral resolution in
order to search for Mg~II absorption at the galaxy redshifts.  The
intital results are surprising.  Although all the galaxies lie within
the canonical 60~\h\ established as the size of Mg~II-absorbing halos,
{\it we find that 50\% (10/20) of the QSO spectra show no Mg~II lines
at the redshifts of the intervening
galaxies}. Figure~\ref{fig_spec_yes} shows some examples of our {\it
MMT} spectra in which the Mg~II absorption was detected, and
Figure~\ref{fig_spec_no} shows cases where no Mg~II lines are apparent
at the expected wavelength. In both figures, the redshift of the
foreground galaxy and the expected wavelengths of the Mg~II doublet
are indicated above the spectrum, and the projected distance between
the galaxy and the sightline are noted at in the lower right corner of
the panel.

Could selection effects explain our finding that the detection rate is
lower than the 100\% expected from earlier studies? Several comments
about this issue can be made. First, the non-absorbing galaxies are no
fainter than the absorbing ones: all have absolute magnitudes in the
range $M_r-5\log h_{70}^{-1} = -20$ to $-23$ (between 0.3 and 5 $L^*$
when compared to the $r$-band luminosity function of galaxies at
$z\sim0.1$ constructed by Blanton~\etal\ 2003).  Second, the detection
of Mg~II is not determined by how close a galaxy is to the QSO
sightline.  Figure~\ref{fig_ew} summarises our results by showing the
rest equivalent width ($W_{\rm r}$) of the Mg~II~$\lambda 2796$
absorption line --- or the $2\sigma$ upper limit --- at the redshift
of the intervening galaxy, plotted against the separation of the
galaxy from the QSO sightline. The non-detections do not simply arise
because a galaxy is further away from the sightline.  Preliminary
analysis shows no clear dependence on galaxy luminosity
either. Finally, the non-detections are not a consequence of
differences in sensitivity of the spectra: all of our QSO spectra
reach similar rest equivalent width limits of $\approx 0.1$~\AA, a
value which is small compared to the Mg~II lines used to define halo
sizes in the previous work. The main difference between our program
and earlier work is whether or not the targets are selected by having
known Mg~II absorption.  It appears that if Mg~II absorption is
present at a particular redshift, there is a high probability that a
galaxy will be found nearby, but the converse does not necessarily
hold.

There are many possible reasons that some galaxies have affiliated
Mg~II absorption and others do not. For example, the detection of
Mg~II absorption could depend on galaxy morphology: perhaps a mixture
of late and early type galaxies, or interacting and non-interacting
galaxies, is responsible for the 50~\% success rate of detections in
our sample.  However, Steidel et al. (1994) found that Mg~II-selected
galaxies have $B-K$ colors indicative of all normal galaxy types,
including ellipticals. High-resolution imaging of these systems would
be very valuable for further investigation of the nature of the
absorbing and non-absorbing galaxies. Another possibility is that the
presence or absence of Mg~II absorption depends on environmental
factors.  For example, galaxies that are embedded in hot intragroup or
intracluster gas might lack associated Mg~II absorption because the
halo gas is highly ionized or has been swept away by ram
pressure. Again, this hypothesis can be tested with follow-up studies,
e.g., deep galaxy redshift surveys over a much larger field centered
on the QSO and X-ray imaging observations.  Ultimately, comparsion
between the properties and environments of absorbing and non-absorbing
galaxies, when significant numbers of both are available, will reveal
the nature and origin of the absorbing gas which, in turn, will make
the technique more useful for the broader goals of galaxy evolution
and cosmology.

\section{Low-Redshift Ly$\alpha$ Forest Systems}

As noted above, Mg~II is not an ideal ion for tracing the distribution
and physical conditions of gas in, around, and (especially) far away
from galaxies. Truly intergalactic gas with low density would likely
be significantly ionized by photoionization from the UV background
from QSOs and AGNs. Intergalactic gas can also have low
metallicity. These ionization and metallicity factors can make Mg~II
undetectable in such gas. Clearly, H~I observations are much more
fundamental. While the hydrogen can also be extremely ionized in the
low-density IGM, H~I Ly$\alpha$ remains detectable because hydrogen is
so abundant.

However, since the longest-wavelength resonance H~I transition is the
Ly$\alpha$ line at 1215.67 \AA , ultraviolet spectroscopy is required
for low-redshift observations and absorber-galaxy correlation
studies. Before the launch of {\it HST}, it was not clear that many
low$-z$ Ly$\alpha$ forest lines would be detected in a typical sight
line. At high redshifts, the number of Ly$\alpha$ clouds per unit
redshift ($dN/dz$) declines rapidly with decreasing redshift, and as
an example, extrapolation of the high$-z$ $dN/dz$ trend to $z = 0$
predicts that one or two Ly$\alpha$ lines with $W_{\rm r} >$ 50 m\AA\
should be found in the sight line to 3C 273.  Consequently, it was a
delightful surprise when the first {\it HST} observations of 3C 273
revealed {\bf ten} Ly$\alpha$ clouds with $W_{\rm r} >$ 50 m\AA\
(Morris et al. 1991). This was a boon for Lya cloud - galaxy studies;
useful samples could evidently be accumlated more rapidly than
expected.  The Ly$\alpha$ $dN/dz$ vs. $z$ has now been measured well
(Penton, Stocke, \& Shull 2004; Williger et al. 2005) and is
theoretically understood: at high-redshift, $dN/dz$(H~I) falls steeply
with $z$ due to decreasing gas density caused by expansion of the
Universe. However, the space density of QSOs also declines
substantially at $z \ll 3$ which, in turn, causes the UV background
intensity and the H~I photoionization rate to decrease markedly at
lower redshifts; the fading of the UV background counteracts the
decreasing density so that the forest does not thin out as rapidly as
expected.  The observed $dN/dz$(H~I) vs. $z$ is also well-matched in
hydrodynamic simulations of cosmological structure growth (e.g.,
Dav\'{e} et al. 1999).

Soon after the first {\it HST} observations, investigations of the
Ly$\alpha$ -- galaxy connection were initiated.  It is worth recalling
that before {\it HST} (and Keck), the prevailing paradigm associated
(high H~I column) metal absorbers with galaxies, and the (lower H~I
column) Ly$\alpha$ forest clouds were thought to be metal-free,
pristine clouds uncorrelated with galaxies (e.g., Sargent et
al. 1980). In this context, initial {\it HST} studies that found
Ly$\alpha$ clouds associated with intervening galaxies in a few cases
(e.g., Bahcall et al. 1992; Spinrad et al. 1993) attracted
considerable interest. Soon thereafter, ground-based echelle
spectroscopy with the Keck 10m telescope showed that metals are
widespread even in low-$N$(H~I) Ly$\alpha$ forest clouds (Tytler et
al. 1995; Cowie et al. 1995) and a major shift in the QSO absorber/IGM
paradigm began. It was apparent that the Ly$\alpha$ clouds are not
randomly distributed with respect to galaxies, and moreover they are
somehow enriched by nucleosynthesis products from stars.

\begin{figure}
\includegraphics[height=7cm]{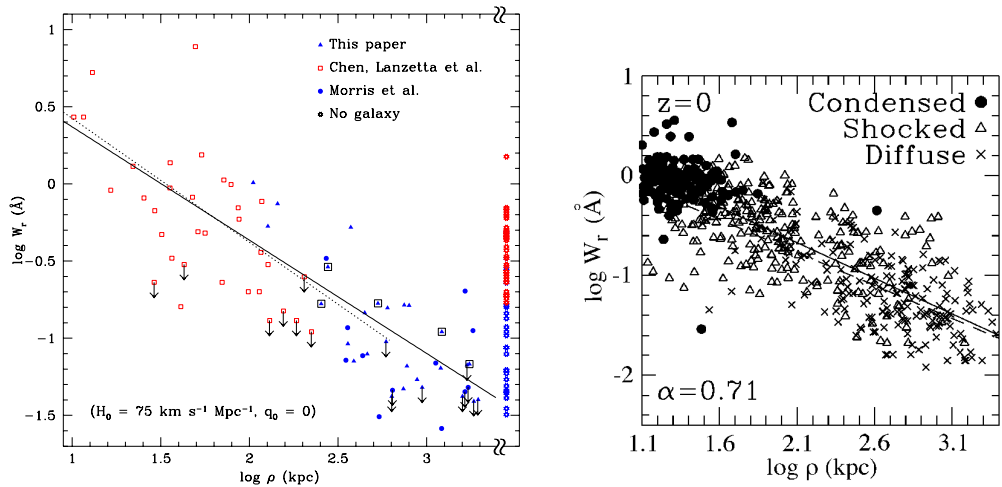}
\caption{({\it Left}) Observed Ly$\alpha$ absorption-line rest
equivalent width vs. impact parameter ($\rho$) of the galaxy with the
lowest $\rho$ at the redshift of the absorber. These observations were
compiled by Tripp et al. (1998) and include the measurements from
Morris et al. (1993) and Chen et al. (1998) as well as the results
from the sight lines studied by Tripp et al. (1998). The open stars on
the right side are Ly$\alpha$ lines that had no known galaxy at the
absorber redshift. The solid and dotted lines show power law fits
($W_{\rm r} \propto \rho ^{\alpha}$) with $\alpha = -0.80 \pm
0.10$. ({\it Right}) Ly$\alpha$ equivalent width vs. $\rho$ from
artificial spectra drawn at random from the hydrodynamic simulations
of the low-redshift IGM from Dav\'{e} et al. (1999). An analogous
power-law fit to the simulated spectra (solid line) yields $\alpha =
-0.71$, in agreement with the observations. In the simulations, some
Ly$\alpha$ lines arise in the bound interstellar gas of condensed
galaxies (filled circles), some in the shock-heated WHIM (open
triangles), and some in the cool, diffuse IGM
(crosses).\label{lyacorrelation}}
\end{figure}

However, as more extensive spectroscopic surveys of galaxy redshifts
near low$-z$ QSO sight lines were executed and analyzed, a debate
about the nature of low$-z$ Ly$\alpha$ lines emerged. One school
argued that the low$-z$ Ly$\alpha$ lines, like Mg~II systems, arise
primarily in large, spherical gaseous halos of {\it individual}
galaxies with near-unity covering factor and $R \approx$ 200 kpc
(Lanzetta et al. 1995; Chen et al. 1998, 2001). This interpretation
was mainly built upon an observed anticorrelation between impact
parameter and Ly$\alpha$ equivalent width, i.e. the lower the
projected galaxy-sight line distance, the greater the Ly$\alpha$
equivalent width (see the left panel of Figure~\ref{lyacorrelation}).
On the other hand, several groups --- using more sensitive data ---
found that the while Ly$\alpha$ lines are significantly correlated
with galaxies, the correlation is not as strong as the galaxy-galaxy
correlation, and indeed some Ly$\alpha$ clouds were found to be
located in galaxy voids (Morris et al. 1993; Stocke et al. 1995;
Tripp, Lu, \& Savage 1998; Impey, Petry, \& Flint 1999).  To reconcile
these discordant results, it was suggested that strong Ly$\alpha$
clouds with $W_{\rm r} >$ 300 m\AA\ --- which dominate the samples of
Lanzetta, Chen et al. --- arise in individual halos, and the
Ly$\alpha$ lines with $W_{\rm r} <$ 200 m\AA\ are less strongly
correlated with galaxies but still are largely located in the same
large-scale dark matter structures as the galaxies.  Using a sample of
QSO - galaxy pairs at very low redshifts ($cz <$ 4000 km s$^{-1}$),
Bowen, Pettini, \& Blades (2002) have also confirmed the $\rho -
W_{\rm r}$ anticorrelation. However, they hypothesize that the
Ly$\alpha$ line strength is correlated with the volume density of
nearby galaxies rather than the impact parameter to a particular
galaxy (see their Figure 19). The basic idea is that as dark matter
potential wells grow deeper, the density of galaxies and the density
of accumulated gas both increase leading to this correlation.

A similar interpretation is favored by theoretical work. Hydrodynamic
cosmological simulations reproduce the impact paramemter - equivalent
width anticorrelation.  Figure~\ref{lyacorrelation} compares the
observed anticorrelation [left panel] to the same anticorrelation
measured from artificial spectra from the cosmological simulation of
Dav\'{e} et al. (1999) [right panel]. The power law fitted to the
observed data ($W_{\rm r} \propto \rho ^{\alpha}$) has the same slope
as the power law from the simulations ($\alpha = -0.80 \pm 0.10$
vs. $\alpha = -0.71$ from the $\Lambda$CDM model of Dav\'{e} et
al. 1999). Using the algorithm of Gelb \& Bertschinger (1994),
Dav\'{e} et al. (1999) have identified galaxies in their simulation,
which in turn they use to determine which Ly$\alpha$ lines originate
in the bound halos of individual galaxies. They find that the impact
paramemter - equivalent width anticorrelation is not strictly due to
absorption arising in particular single galaxies. The filled circles
in the right panel of Figure~\ref{lyacorrelation} arise in bound halos
of ``condensed'' individual galaxies; these cases are mostly at $\rho
<$ 50 \h . Thus, Ly$\alpha$ absorbers at larger impact parameters
are not bound to individual galaxies but nevertheless are located in
the same large-scale structures as the nearby galaxies.
Interestingly, Dav\'{e} et al. find that some of the Ly$\alpha$ lines
near galaxies come from shock-heated gas at $T > 10^{5}$ K, which has
some implications for the missing baryons problem (see \S
\ref{whimsec}). Others are from ``diffuse'' clouds, photoionized and
relatively cool gas (at $T \sim 10^{4}$ K), which are most analogous
to the gas clouds that dominate the high$-z$ Ly$\alpha$ forest.

There remain many open questions about the nature of low$-$z
Ly$\alpha$ clouds. Existing studies suffer from a variety of selection
effects (see, e.g., Tripp et al. 1998). A particularly serious problem
with Ly$\alpha$ observations carried out with first-generation {\it
HST} spectrographs is that many (but not all) of the spectra have low
spectral resolution, i.e., FWHM $\geq$ 150 km s$^{1}$. One of the
original {\it HST} instruments, the Goddard High-Resolution
Spectrograph (GHRS), was capable of recording high-resolution spectra
(e.g., Morris et al. 1991; Cardelli et al. 1991, 1993; Bowen et
al. 1995). However, the one-dimensional detectors of the GHRS severely
limited the wavelength range of individual exposures obtained in
high-resolution modes, and to obtain broad wavelength coverage
required prohibitive telescope time.  The installation of STIS, which
used two-dimensionsal detectors and therefore could observe wide
wavelength ranges at high resolution in single exposures, provided a
great leap in capability.

\begin{figure}
\includegraphics[height=10cm]{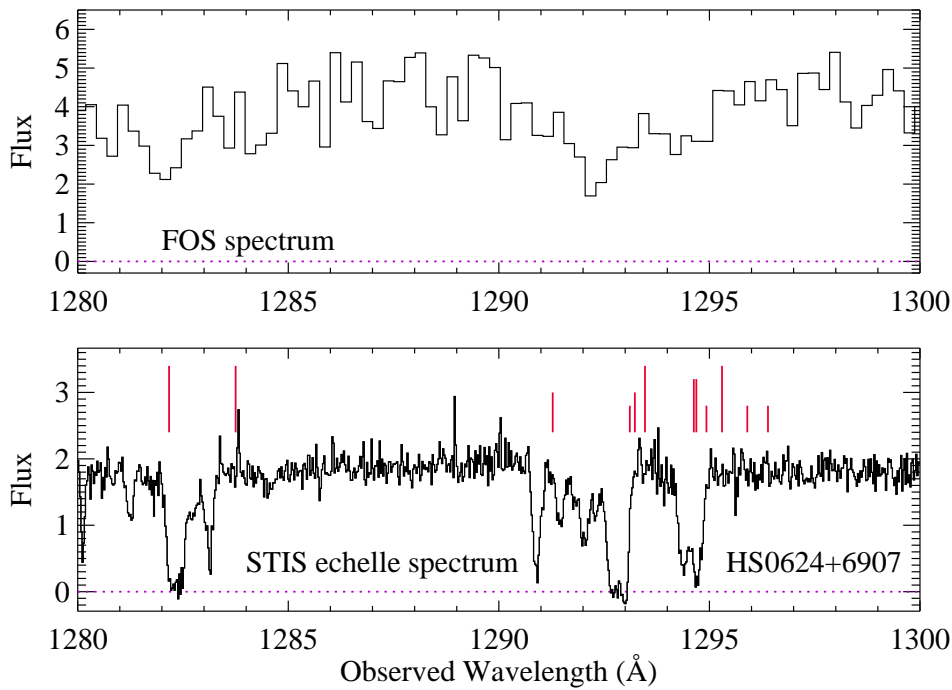}
\caption{Comparison of a portion of the Faint Object Spectrograph
observation of HS0624+6907 (upper panel) from Bechtold et al. (2002)
to the same portion of the HS0624+6907 spectrum observed with the
E140M echelle mode of STIS (lower panel) by Aracil et al. 2005. The
FOS spectral resolution is $\sim 150$ km s$^{-1}$ vs. the 7 km s$^{1}$
resolution provided by the STIS echelle. The bars above the spectrum
indicate the redshifts of galaxies near the sight line, and the length
of the bar reflects the proximity of the galaxy in projection (longer
bars are closer to the sight line); see Aracil et al. for details
about these galaxies.\label{fosvsstis}}
\end{figure}

The importance of high resolution is illustrated in
Figure~\ref{fosvsstis}. This figure compares the same portion of the
spectrum of the low$-z$ QSO HS0624+6907 observed with the {\it HST}
Faint Object Spectrograph (FOS, FWHM $= 150$ km s$^{-1}$, upper panel)
from Bechtold et al. (2002, see also Jannuzi et al. 1998) and the
E140M echelle mode of STIS (FWHM = 7 km s$^{-1}$, lower panel) from
Aracil et al. (2005). It is immediately obvious from this comparison
that crucial information is lost in the low-resolution spectrum.  We
notice several problems: (1) The FOS mainly detects strongly saturated
lines, which makes accurate column density measurements very difficult
or impossible. (2) In this case, there is a substantial error in the
FOS wavelength zero point (the FOS is shifted by by $\sim$80 km
s$^{-1}$ with respect to the STIS spectrum, which has an accurate zero
point. This is only half of an FOS resolution element, but
nevertheless such errors can cause confusion if velocity
differences are used to match a particular Ly$\alpha$ absorber with a
particular galaxy (at Ly$\alpha$ redshifts, there are often multiple
galaxies with comparable $\rho$ and $\Delta v$ values) or if galaxy
velocity curves are compared to the line centroids. (3) The kinematics
and component structure revealed by high resolution substantially
improve our understanding of the absorption. Consider the cluster of
Ly$\alpha$ lines evident between 1290 and 1295 \AA\ in
Figure~\ref{fosvsstis}. We find that 13 H~I components are spread over
this 1000 km s$^{-1}$ interval.  These kinematics cannot be explained
by rotation of a galaxy disk or halo.  Even tidal stripping is
unlikely to produce such a velocity spread.  Instead, this absorption
cluster is likely intragroup gas that is not bound to a particular
galaxy.  On the other hand, the FOS data can only indicate the
presence of one or two components in this wavelength range, which
could be erroneously interpreted as simple, ordinary halo gas
associated with the two nearest galaxies (the two longest bars in
Figure~\ref{fosvsstis}). We expect that the current archive of
high-resolution STIS spectra will significantly advance our knowledge
of the Ly$\alpha$-galaxy connection in upcoming years (as shown in,
e.g., Bowen et al. 2002; Penton et al. 2004; Richter et al. 2004;
Sembach et al. 2004; Williger et al. 2005; Aracil et al. 2005)

\section{Lyman Limit, Sub-damped, and Damped Ly$\alpha$ Absorbers}

Many papers in these proceedings discuss the higher-$N$(H~I)
absorption systems, i.e., the Lyman limit ($10^{16.5} < N$(H~I) $<
10^{19}$ cm$^{-2}$), sub-damped Ly$\alpha$ ($10^{19.0} < N$(H~I) $<
10^{20.3}$ cm$^{-2}$), and damped Ly$\alpha$ ($N$(H~I) $> 10^{20.3}$
cm$^{-2}$) absorption systems and their relationships with galaxies.
Furthermore, most of the Mg~II absorbers discussed in \S \ref{secmg2}
are actually in these high-$N$(H~I) categories; most Mg~II systems
have log $N$(H~I) $>$ 16.  However, we would like to add a few
additional comments about these types of absorbers.

In the nearby universe, relatively large clouds of intergalactic H~I
have been known from 21 cm emission studies for many years; these
would be LL, sub-DLA, and DLA absorbers if a QSO were behind
them. Some well-known examples include H~I 1225+01 (Giovanelli \&
Haynes 1989), the ``Leo Ring'' in the M96 group (Schneider 1989),
HIPASS J1712-64 (Kilborn et al. 2000), and most recently,
``VIRGOHI21'' (Minchin et al. 2005).  These clouds have attracted
considerable interest, but it is often not clear what role they play
in galaxy evolution and cosmology. They could be the leftover building
blocks of galaxy formation, i.e., the small dark matter and gas halos
that merge to form large galaxies in hierarchical formation scenarios
(e.g., Blitz et al. 1999), but many other interpretations remain
viable, e.g., they could be debris from tidal or ram-pressure
stripping. Recently Minchin et al. have suggested that VIRGOHI21 is
the first discovery of a ``dark galaxy'', a large dark matter halo
($\geq 10^{11}$ M$_{\odot}$) that for some reason has failed to form
detectable stars. If correct, this is an important finding for
cosmology and galaxy evolution. However, Oosterloo \& van Gorkom
(2005) have argued that VIRGOHI21 might instead be material
dynamically stripped from the nearby galaxy NGC4254.

The nature and importance of these H~I clouds are poorly
understood partly because 21 cm emission provides no information about
the chemical enrichment and star-formation history of the
objects. Galaxies are usually found in or near the clouds, but whether
the gas is entering or leaving the galaxy (or neither) is
unclear. Chemical enrichment measurements can elucidate the nature of
{\sc H i} clouds: tidal or ram-pressure stripped gas should be
enriched to the metallicity level of the source galaxy. Small and old
dark matter halos, on the other hand, are expected to be metal poor.

Hence, there is need to combine information from QSO absorption
measurements with other techniques such as 21 cm emission mapping. We
have recently discovered an intergalactic H~I cloud similar to
VIRGOHI21, but uniquely our cloud was discovered {\it in Ly$\alpha$
absorption} in an ultraviolet {\it HST} spectrum of a background QSO
(Tripp et al. 2005). The sight line to this QSO (PG1216+069, $z_{\rm
QSO}$ = 0.331) passes through the outskirts of the Virgo cluster close
to the X-ray bright NGC4261 galaxy group.  We obtained a
high-resolution (7 km s$^{-1}$ FWHM) UV echelle spectrum of
PG1216+069, and this revealed damped Ly$\alpha$ absorption at the
Virgo redshift $z_{\rm abs}$ = 0.00632 (see Figure 2 in Tripp et
al. 2005) with log $N$({\sc H i}) = 19.32$\pm$0.03. Metal absorption
lines in the echelle spectrum robustly indicate that the metallicity
is quite low: [O/H] = $-1.60^{+0.09}_{-0.11}$. Moreover, nitrogen is
significantly underbundant, which suggests that intermediate-mass stars have
not contributed substantially to the metal enrichment. Evidently, this
is a relatively primitive damped Ly$\alpha$ absorber. We have searched
for associated galaxies using {\it HST} imaging and spectroscopic
redshift surveys, but we are unable to find a nearby galaxy. The
closest known galaxies are at substantial projected distances from the
sight line (see Tripp et al. 2005).  Clouds like this one appear to be
not uncommon in the spectra of low$-z$ QSOs.  Additional information
from techniques like 21 cm emission observations could provide
valuable information such as the physical dimensions of the cloud, its
H~I mass and dynamical mass, and its morphology.  Conversely,
deployment of a more a new, substantially more sensitive,
high-resolution UV spectrograph (e.g., Sembach et al. 2005) could
enable studies of objects that have already been mapped in, e.g. 21
cm, because faint QSOs are known behind many interesting targets, but
they are too faint for current UV facilities.

Like the absorbers discussed in previous sections, the high-$N$(H~I)
systems probably have a variety of origins.  The Giovanelli \& Haynes
cloud spans $\sim 200$ kpc in its long dimension, for example, and the
properties of some QSO absorbers imply similar sizes (e.g., Savage et
al. 2002,2005).  On the other hand, some high-$N$(H~I) systems have
properties that imply vastly smaller sizes (e.g., Tripp et al. 2002;
Rigby, Charlton, \& Churchill 2002), sizes of order a few tens to few
hundreds of parsecs.  Stocke et al. (2004) have found a dwarf
poststarburst galaxy near the small cloud studied by Tripp et
al. (2002), and they have suggested that the absorber is an ejecta
fragment in a supernova-driven wind.  The dwarf galaxy is relatively
faint ($M_{B} = -13.9$) and has a projected distance of 71 \h from
the sight line.  In order to search for similar galaxies near other
QSO absorption systems, very deep observations will be needed.  This
poses a serious challenge for absorber-galaxy studies.

\begin{figure}
\includegraphics[height=10.5cm]{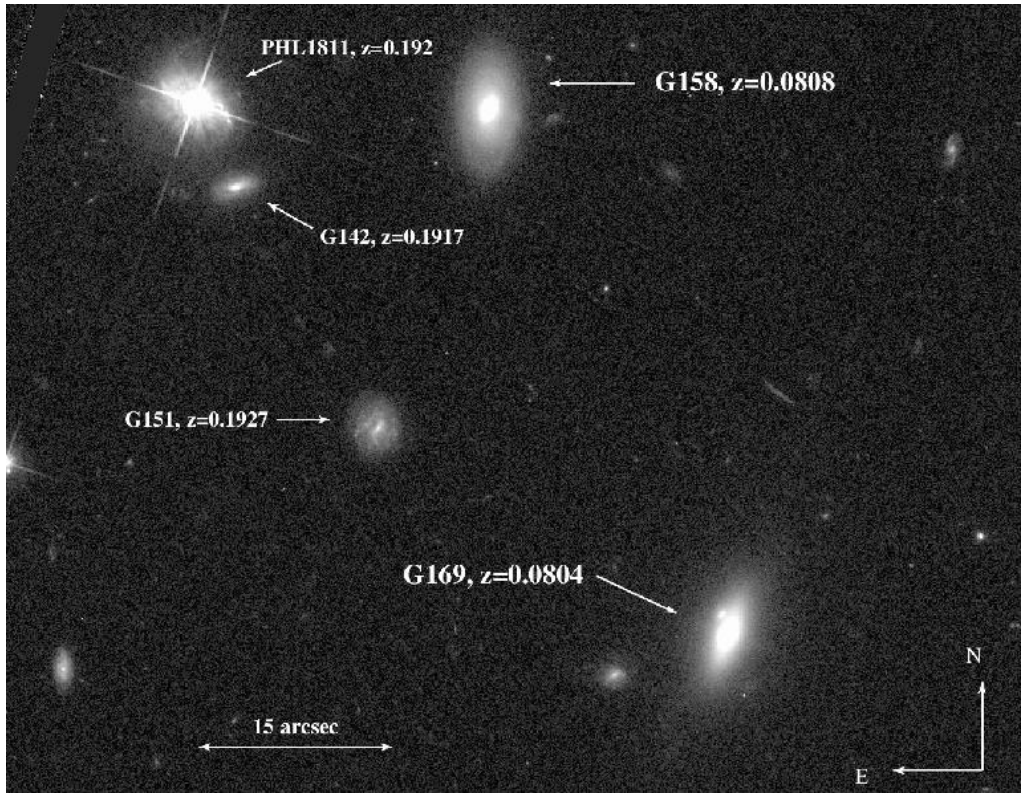}
\caption{Portion of an image of the field near the bright QSO PHL1811
($z_{\rm QSO}$ = 0.192) from Jenkins et al. (2005). Galaxies in the
image with known redshifts are labeled with a name and redshift from
Jenkins et al. (2003). The image was obtained with the Advanced Camera
for Surveys on board {\it HST}, and spectroscopic galaxy redshifts
were measured with the Double Imaging Spectrograph on the APO 3.5m
telescope (see Jenkins et al. 2003 for details).\label{phl1811image}}
\end{figure}

Some high-$N$(H~I) systems have shown surprising abundances patterns.
For example, we have recently completed a detailed investigation of a
Lyman limit absorber at $z=0.0809$ towards the bright QSO
PHL~1811 (Jenkins et al. 2005). This LL system shows a rather unusual
abundance pattern with a high oxygen metallicity but an extrememly low
nitrogen abundance.  Underabundances of nitrogen are expected (and
often observed) in low metallicity systems because nitrogen is
synthesized in longer-lived intermediate-mass stars and moreover its
synthesis depends on the initial metallicity of the gas (see, e.g.,
Henry, Edmunds, \& K\"{o}ppen 2000; Prochaska et al. 2002; Pilyugin et
al. 2003, and references therein).  In these cases, the nitrogen
production is lagging behind the rapid $\alpha-$element enrichment
from type II SNe.  However, it is surprising to find that nitrogen is
underabundant in high-metallicity gas.  Possibly this abundance
pattern is due to a massive burst of star formation that enriched and
blew out the interstellar gas before intermediate-mass stars were able to
substantially enrich the gas with nitrogen.  High-resolution imaging
of the PHL1811 field obtained with the {\it HST} Advanced Camera for
Surveys (see Figure~\ref{phl1811image}) has revealed two S0 galaxies
at the Lyman limit redshift near the sight line (see Jenkins et
al. 2005 for details), so ram-pressure stripping or tidal interactions
may also play a role in the origin of this absorber.

\section{O~VI Absorbers, Broad Ly$\alpha$ Lines, and the WHIM\label{whimsec}}

We close this article with a brief review of O~VI absorption systems
and their connections to galaxies. These absorbers appear to harbor a
significant fraction of the baryons at the present epoch (Tripp,
Savage, \& Jenkins 2000; Tripp \& Savage 2000; Savage et al. 2002,
2005; Sembach et al. 2004; Richter et al. 2004; Danforth \& Shull
2005).  Again this observation jibes well with the general results of
cosmological simulations because the O~VI ion fraction peaks at $T
\approx 10^{5.5}$ K in collisional ionization equilibrium, and
therefore these absorbers have the potential to reveal the so-called
``warm-hot'' intergalactic medium (WHIM) at $T = 10^{5} - 10^{7}$ K
that is predicted by the cosmological simulations to contain 30-50\%
of the baryons at the present epoch (Cen \& Ostriker 1999; Dav\'{e} et
al. 1999).

\begin{figure}
\includegraphics[height=8cm]{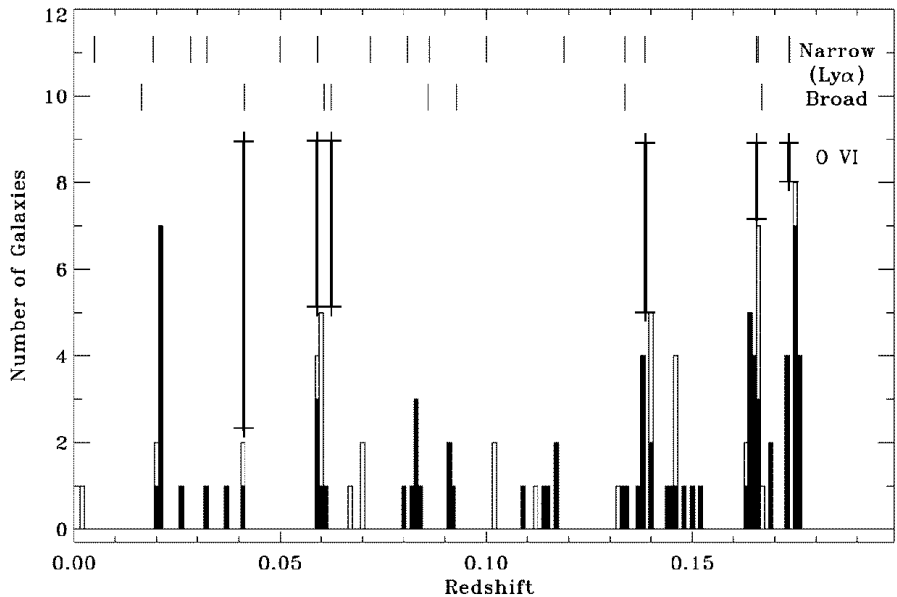}
\caption{Comparison (from Sembach et al. 2004) of the galaxy redshift
distribution in the $\sim 1^{\circ}$ field centered on PG1116+215 to
the redshifts of narrow and broad Ly$\alpha$ absorption lines (upper
two rows of tick marks) and O~VI lines (long tick marks with cross
bars). The galaxies are binned in $\Delta z$ = 0.001 intervals. The
open histogram shows the redshifts of galaxies within 50' of the
sightline while the filled histogram indicates galaxies within
30'.\label{pg1116reds}}
\end{figure}

\begin{figure}
\includegraphics{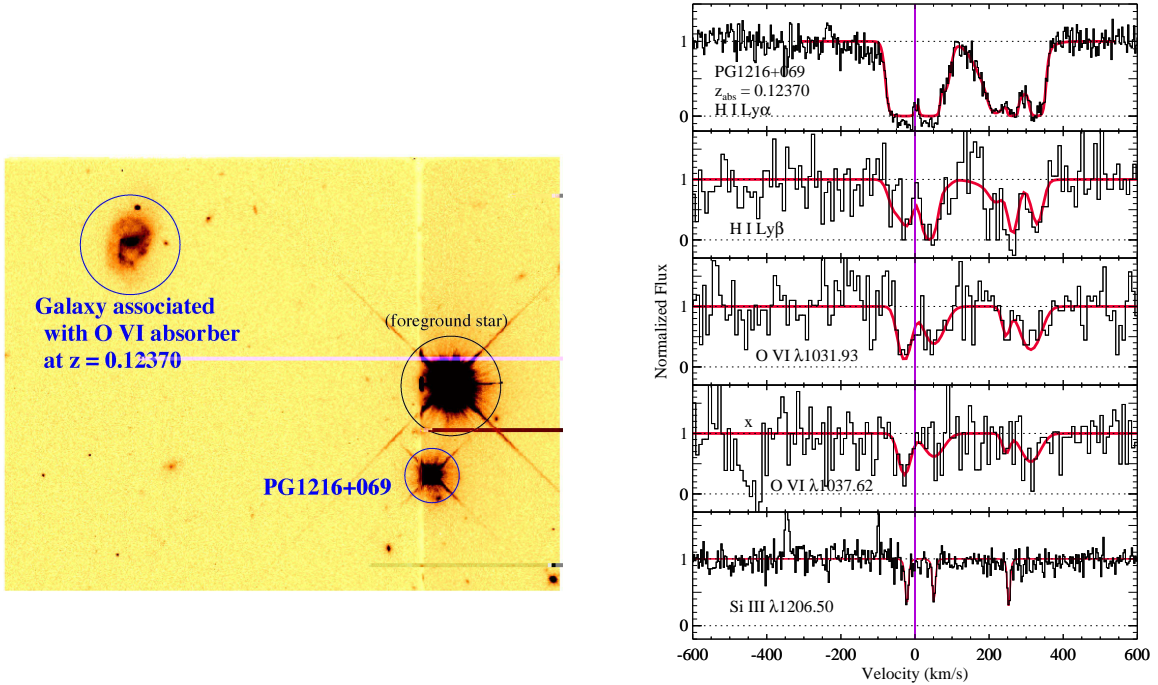}
\caption{O~VI absorption-line system (right panels) associated with a
spiral galaxy at a projected distance of 88 \h kpc from the sight line
to QSO PG1216+069 (left panel). Voigt profile fits are overplotted on
the observed data in the right panels, which are plotted in the
velocity frame of the absorber ($v = 0$ km s$^{-1}$ at $z = 0.12370$.)
The image is a portion of the {\it HST}/WFPC2 data obtained by Chen et
al. (2001). \label{oviexample}}
\end{figure}

Studies of O~VI - galaxy connections are in their infancy.  However,
galaxy redshift surveys have been conducted in the fields of several
QSOs that show intervening O~VI absorbers (e.g., H1821+643,
PG1116+215, PG0953+415, PKS0405-123), and in these directions, the
O~VI absorbers appear to be strongly correlated with galaxy
structures. Figure~\ref{pg1116reds} compares the distribution of galaxies
as a function of redshift in the field of PG1116+215 (histograms) to
the distribution of O~VI absorbers (long thick lines with cross
bars). At a glance, the O~VI absorbers appear to mostly follow the
galaxies, and this is confirmed by statistical tests: Sembach et
al. (2004) show that the probability that the O~VI systems are
randomly distributed with respect to the galaxies in the field is
extremely small. Case studies of specific O~VI systems show that O~VI
absorption arises in a variety of locations (see, e.g., Tripp \&
Savage 2000; Chen \& Prochaska 2000; Tripp et al. 2001; Sembach et
al. 2001; Savage et al. 2002; Shull, Tumlinson, \& Giroux 2003;
Tumlinson et al. 2005) ranging from modest overdensity regions of
large-scale structures to galaxy groups to relatively dense regions in
the Virgo cluster.  Figure~\ref{oviexample} shows an example of an
O~VI system associated with a spiral galaxy at a projected distance of
88 kpc.  The absorption line profiles in this case have a symmetric
shape reminiscent of the Mg~II absorbers presented by Bond et
al. (2001), and this absorption could arise in a similar outflow
scenario. Three components are significantly detected in the Si~III
$\lambda$1206.5 transition.  This may seem surprising since
supernova-driven, outflowing hot gas might be expected to ionize
silicon to the degree where very little Si~III is present.  However,
optical observations of outflows from nearby superwind and
ultraluminous infrared galaxies show that even Na~I (which is much
more easily ionized than Si~III) survives and is accelerated to
substantial velocities in such outflows (e.g., Heckman et al. 2000;
Rupke, Veilleux, \& Sanders 2002; Martin 2005), so some aspect of the
outflow process can accelerate material considerably without severely
ionizing it.  The narrow Si~III components that are associated with
the O~VI lines shown in Figure~\ref{oviexample} may be analogous
lower-ionization clouds in an outflow.  Indeed, in the hydrodynamic
models of a superwind outflow presented by Heckman et al. (2002), O~VI
is predicted to exist in interface regions between low-ionization
clumps and the hotter X-ray emitting gas.

A persistent problem with the use of O~VI absorption lines as a probe
of the warm-hot intergalactic medium is that often the O~VI doublet
and the H~I Ly$\alpha$ lines are the only detected transitions in a
particular absorber.  With only these transitions, the metallicity of
the gas can only be loosely constrained, and this introduces
considerable uncertainty into the baryonic content of the absorber.
Another means to search for WHIM gas is to look for very broad
Ly$\alpha$ lines.  As shown in Figure~\ref{lyacorrelation}, the
cosmological simulations predict that a reasonable fraction of
Ly$\alpha$ lines in the vicinity of a galaxy should occur in
shock-heated gas (open triangles in
Figure~\ref{lyacorrelation}). Consulting Dav\'{e} et al. (1999), we
find that these shocked Ly$\alpha$ lines are predicted to have gas
temperatures ranging from $10^{5}$ to $10^{6}$ K and therefore should
be quite broad. Tripp et al. (2001) reported a broad Ly$\alpha$ line
associated with an O~VI absorber that arises in a small group of
galaxies; this system is a strong candidate for WHIM gas.  Richter et
al. (2004) and Sembach et al. (2004) have recently identified a number
of broad Ly$\alpha$ lines in high-resolution STIS spectra of the QSOs
PG1259+593 and PG1116+215.  Estimates of the baryonic content of these
broad Ly$\alpha$ candidates suggests that their baryonic content is
comparable to that of the O~VI systems.  Again, a better UV
spectrograph would likely lead to real progress on understanding broad
Ly$\alpha$ candidates and their relationship to the WHIM: the current
data suffer from limited signal-to-noise ratios, and this can cause
blends, for example, to mimic smooth and broad Ly$\alpha$ features.
However, a more sensitive UV spectrograph could be used to obtain much
higher S/N data.  With higher S/N, the shape of a Ly$\alpha$ profile
can be carefully inspected to check whether the absorption is truly a
single broad Gaussian or if it shows asymmetries and signs of blended
component structure.  The Cosmic Origins Spectrograph, which is now
built and awaiting installation in the {\it Hubble Space Telescope},
is ideal for this topic and indeed would be very useful for all of the
topics presented in this brief review.

\begin{acknowledgments}
This work was supported by NASA through the Long-Term Space
Astrophysics grant NNG 04GG73G.
\end{acknowledgments}



\begin{thebibliography}{}

\bibitem[Aguirre et al. (2001)]{aguirre01} {Aguirre, A., Hernquist,
L., Schaye, J., Weinberg, D.H., Katz, N., \& Gardner, J.} 2001,
\textit{ApJ} 560, 599 

\bibitem[Aracil et al. (2005)]{aracil05} {Aracil, B., Tripp, T. M.,
Bowen, D. V., Prochaska, J. X., Chen, H.-W., \& Frye, B. L.} 2005, \textit{MNRAS},
submitted

\bibitem[Bahcall et al. (1992)]{bah92} {Bahcall, J. N., Jannuzi,
B. T., Schneider, D. P., Hartig, G. F., \& Green, R. F.} 1992,
\textit{ApJ}, 397, 68

\bibitem[Bechtold et al. (2002)]{bech02} {Bechtold, J., Dobrzcki, A.,
Wilden, B., Morita, M., Scott, J., Dobrzcka, D., Tran, K.-V., \&
Aldcroft, T. L.} 2002, \textit{ApJS}, 140, 143

\bibitem[Ben\'{i}tez et al. (2004)]{ben04} {Ben\'{i}tez, N. et al.}
2004, \textit{ApJS}, 150, 1

\bibitem[Bergeron (1986)]{berge86} {Bergeron, J.} 1986,
\textit{A\&A}, 155, L8

\bibitem[Bergeron \& Boiss\'{e} (1991)]{berge91} {Bergeron, J. \&
Boiss\'{e}} 1991, \textit{A\&A}, 243, 344

\bibitem[Blanton et al. (2003)]{blanton} {Blanton, M.R. et al.} 2003,
\textit{ApJ} 592, 819

\bibitem[Blitz et al. (1999)]{blitz} {Blitz, L., Spergel, D. N., Teuben, P. J., Hartmann, D., \& Burton, W. B.} 1999, \textit{ApJ}, 514, 818

\bibitem[Bond et al. (2001)]{bond} {Bond, N. A., Chuchill, C. W.,
Charlton, J. C., \& Vogt, S. S.} 2001, \textit{ApJ}, 562, 641

\bibitem[Bowen et al. (1995)]{bowen95} {Bowen, D. V., Blades, J. C.,
\& Pettini, M.} 1995, \textit{ApJ}, 448, 634

\bibitem[Bowen et al. (2002)]{bowen02} {Bowen, D. V., Pettini, M., \&
Blades, J. C.} 2002, \textit{ApJ}, 580, 169

\bibitem[Cardelli et al. (1993)]{card93} {Cardelli, J. A., Federeman,
S. R., Lambert, D. L., \& Theodosiou, C. E.} 1993, \textit{Ap.J.},
416, L41

\bibitem[Cardelli et al. (1991)]{card91} {Cardelli, J. A., Savage,
B. D., \& Ebbets, D. C.} 1991, \textit{ApJ}, 383, L23

\bibitem[Cen \& Ostriker (1999)]{cen99} {Cen, R. \&
Ostriker,~J.~P.} 1999, \textit{ApJ}, 514, 1

\bibitem[Charlton \& Churchill (1998)]{char98} {Charlton, J. C., \&
Churchill, C. W.} 1998, \textit{ApJ}, 499, 181

\bibitem[Chen et al. (1998)]{chen98} {Chen, H.-W., Lanzetta, K. M.,
Webb, J. K., \& Barcons, X.} 1998, \textit{ApJ}, 498, 77

\bibitem[Chen et al. (2001)]{chen01} {Chen, H.-W., Lanzetta, K. M.,
Webb, J. K., \& Barcons, X.} 2001, \textit{ApJ}, 559, 654

\bibitem[Chen \& Prochaska (2000)]{chen00} {Chen, H.-W., \& Prochaska,
J. X.} 2000, \textit{ApJ}, 543, L9

\bibitem[Churchill, Kacprzak, \& Steidel (2005)]{church05} {Churchill,
C. W., Kacprzak, G. G., \& Steidel, C. C.} 2005, \textit{these proceedings}

\bibitem[Churchill, Steidel, \& Vogt (1996)]{church94} {Churchill,
C. W., Steidel, C. C., \& Vogt, S. S.} 1996, \textit{ApJ}, 471, 164

\bibitem[Colless et al. (2001)]{coll01} {Colless, M., et al.} 2001,
\textit{MNRAS}, 328, 1039

\bibitem[Cowie et al. (1995)]{cow95} {Cowie, L. L., Songaila, A., Kim,
T.-S., \& Hu, E. M.} 1995, \textit{AJ}, 109, 1522

\bibitem[Cristiani (1987)]{cristiani87} {Cristiani, S.} 1987,
\textit{A\&A}, 175, L1

\bibitem[Danforth \& Shull (2005)]{dan05} {Danforth, C. W., \& Shull,
J. M.} 2005, \textit{ApJ}, 624, 555

\bibitem[Dav\'{e} et al. (1999)]{dave99} {Dav\'{e}, R., Hernquist, L.,
Katz, N., \& Weinberg, D.} 1999, \textit{ApJ}, 511, 521

\bibitem[Ellison, Mallen-Ornelas, \& Sawicky (2003)]{ellison03}
{Ellison, S. L., Mallen-Ornelas, G., \& Sawicky, M.} 2003,
\textit{ApJ}, 589, 709

\bibitem[Gelb \& Bertschinger (1994)]{gelb94} {Gelb, J. M., \&
Bertschinger, E.} 1994, \textit{ApJ}, 436, 467

\bibitem[Giovanelli \& Haynes (1989)]{gio89} {Giovanelli, R., \&
Haynes, M. P.} 1989, \textit{ApJ}, 346, L5

\bibitem[Heckman et al. (2000)]{heck00} {Heckman, T. M., Lehnert, M. D.,
Strickland, D. K., \& Armus, L.} 2000, \textit{ApJS}, 129, 493

\bibitem[Heckman et al. (2002)]{heck02} {Heckman, T. M., Norman,
C. A., Strickland, D. K., \& Sembach, K. R.} 2002, \textit{ApJ}, 577,
691

\bibitem[Henry, Edmunds, \& K\"{o}ppen (2000)]{hen00} {Henry,
R. B. C., Edmunds, M. G., \& K\"{o}ppen, J.} 2000, \textit{ApJ}, 541,
660

\bibitem[Hobbs et al. (1993)]{hobb93} {Hobbs, L. M., Welty, D. E.,
Morton, D. C., Spitzer, L., \& York, D. G.} 1993, \textit{ApJ}, 411,
750

\bibitem[Impey, Petry, \& Flint (1999)]{imp99} {Impey, C. D., Petry,
C. E., \& Flint, K. P.} 1999, \textit{ApJ}, 524, 536

\bibitem[Jannuzi et al. (1998)]{jan98} {Jannuzi, B. T. et al.} 1998,
\textit{ApJS}, 118, 1

\bibitem[Jenkins et al. (2005)]{jenk05} {Jenkins, E.B., Bowen, D.V.,
Tripp, T.M. \& Sembach, K.R.} 2005, \textit{ApJ}, 623, 767

\bibitem[Jenkins et al. (2003)]{jenk03} {Jenkins, E. B., Bowen, D.V.,
Tripp, T.M., Sembach, K. R., Leighly, K. M., Halpern, J. P., \&
Lauroesch, J. T.} 2003, \textit{AJ}, 125, 2824

\bibitem[Kilborn et al. (2000)]{kilb00} {Kilborn, V. A. et al.} 2000,
\textit{AJ}, 120, 1342

\bibitem[Lanzetta et al. (1995)]{lanz95} {Lanzetta, K. M., Bowen,
D. V., Tytler, D., \& Webb, J. K.} 1995, \textit{ApJ}, 442, 538

\bibitem[Le Brun et al. (1997)]{lebrun97} {Le Brun, V., Bergeron, J.,
Boisse, P. \& Deharveng, J.M.} 1997, \textit{A\&A}, 321, 733

\bibitem[Maller et al. (2003)]{mall03} {Maller, A. H., McIntosh,
D. H., Katz, N., \& Weinberg, M. D.} 2003, \textit{ApJ}, 598, L1

\bibitem[Martin (2005)]{martin05} {Martin, C. L.} 2005, \textit{ApJ},
621, 227

\bibitem[Minchin et al. (2005)]{minch} {Minchin, R. et al.} 2005,
\textit{ApJ}, 622, L21

\bibitem[Morris et al. (1993)]{morris93} {Morris, S. L., Weymann,
R. J., Dressler, A., McCarthy, P. J., Smith, B. A., Terrile, R. J.,
Giovanelli, R., \& Irwin, M. 1993} \textit{ApJ}, 419, 524

\bibitem[Morris et al. (1991)]{morris91} {Morris, S. L., Weymann,
R. J., Savage, B. D., \& Gilliland, R. L.} 1991, \textit{ApJ}, 377,
L21

\bibitem[Oosterloo \& van Gorkom]{oost} {Oosterloo, T. \& van Gorkom,
J.} 2005, \textit{A\&A}, 437, L19

\bibitem[Penton, Stocke, \& Shull (2004)]{pen04} {Penton, S. V.,
Stocke, J. T., \& Shull, J. M.} 2004, \textit{ApJS}, 152, 29

\bibitem[Pilyugin, Thuan, \& V\'{i}lchez (2003)]{pily03} {Pilyugin,
L. S., Thuan, T. X., \& V\'{i}lchez, J. M.} 2003, \textit{A\&A}, 397,
487

\bibitem[Prochaska et al. (2002)]{proch02} {Prochaska, J. X., Henry,
R. B. C., O'Meara, J. M., Tytler, D., Wolfe, A. M., Kirkman, D.,
Lubin, D., \& Suzuki, N.} 2002, \textit{PASP}, 114, 933

\bibitem[Prochaska et al. (2003)]{proch03} {Prochaska, J. X., Howk,
J. C., \& Wolfe, A. M.} 2003, \textit{Nature}, 423, 57

\bibitem[Richter et al. (2004)]{rich04} {Richter, P., Savage, B. D.,
Tripp, T. M., \& Sembach, K. R.} 2004, \textit{ApJS}, 153, 165

\bibitem[Rigby, Charlton, \& Churchill (2002)]{rig02} {Rigby, J. R.,
Charlton, J. C., \& Churchill, C. W.} 2002, \textit{ApJ}, 565, 743

\bibitem[Rupke, Veilleux, \& Sanders (2002)]{rup02} {Rupke, D. S.,
Veilleux, S., \& Sanders, D. B.} 2002, \textit{ApJ}, 570, 588

\bibitem[Sargent et al. (1980)]{sar80} {Sargent, W. L. W., Young,
P. J., Boksenberg, A., \& Tytler, D.} 1980, \textit{ApJS}, 42, 41

\bibitem[Savage \etal (2005a)]{sav05a} {Savage, B. D., Lehner, N.,
Wakker, B. P., Sembach, K. R., \& Tripp, T. M.} 2005, \textit{ApJ},
626, 776

\bibitem[Savage et al. (2002)]{sav02} {Savage, B. D., Sembach, K. R.,
Tripp, T. M., \& Richter, P.} 2002, \textit{ApJ}, 564, 631

\bibitem[Savage et al. (2000)]{sav00} {Savage, B. D., et al.} 2000,
{\textit ApJS}, 129, 563

\bibitem[Schneider (1989)]{sch89} {Schneider, S. E.} 1989,
\textit{ApJ}, 343, 94

\bibitem[Sembach et al. (2001)]{sem01} {Sembach, K. R., Howk, J. C.,
Savage, B. D., Shull, J. M., \& Oegerle, W. R.} 2001, \textit{ApJ},
561, 573

\bibitem[Sembach et al. (2004)]{sem04} {Sembach, K. R., Tripp, T. M.,
Savage, B. D., \& Richter, P.} 2004, \textit{ApJS}, 155, 351

\bibitem[Sembach et al. (2005)]{sem05} {Sembach, K. R. et al.} 2005,
\textit{The Baryonic Structure Probe: Characterizing the Cosmic Web of
Matter Through Ultraviolet Spectroscopy}, an Origins Probe concept
study submitted to NASA, May 13, 2005

\bibitem[Shull, Tumlinson, \& Giroux (2003)]{shull03} {Shull, J. M.,
Tumlinson, J., \& Giroux, M. L.} 2003, \textit{ApJ}, 594, 107

\bibitem[Spinrad et al. (1993)]{spin93} {Spinrad, H. et al.} 1993,
\textit{AJ}, 106, 1

\bibitem[Steidel (1994)]{steidel94} {Steidel, C.C.} 1994, in G. Meylan
(ed), \textit{QSO Absorption Lines, ESO Astrophys. Symp.}, p. \ 139

\bibitem[Steidel et al. (1994)]{sdp94} {Steidel, C.C., Dickinson, M. \&
Persson, S.E.} 1994, \textit{ApJ} 437, L75

\bibitem[Steidel et al. (2002)]{steidel02} {Steidel, C.C., Kollmeier,
J. A., Shapley, A. E., Churchill, C. W., Dickenson, M., \& Pettini,
M.} 2002, {\textit ApJ}, 570, 526

\bibitem[Stocke et al. (2004)]{stocke04} {Stocke, J. T., Keeney,
B. A., McLin, K. M., Rosenberg, J. L., Weymann, R. J., \& Giroux,
M. L.} 2004, \textit{ApJ}, 609, 94

\bibitem[Stocke et al. (1995)]{stocke95} {Stocke, J. T., Shull, J. M.,
Penton, S., Donahue, M., \& Carilli, C.} 1995, \textit{ApJ}, 451, 24

\bibitem[Stoughton et al. (2002)]{stoughton02} {Stoughton, C., et al.}
2002, \textit{AJ}, 123, 485

\bibitem[Tripp et al. (2001)]{tripp01} {Tripp, T. M., Giroux, M. L.,
Stocke, J. T., Tumlinson, J., \& Oegerle, W. R.} 2001, \textit{ApJ},
563, 724

\bibitem[Tripp et al. (2005)]{tripp05} {Tripp, T. M., Jenkins, E. B.,
Bowen, D. V., Prochaska, J. X., Aracil, B., \& Ganguly, R.} 2005,
\textit{ApJ}, 619, 714

\bibitem[Tripp et al. (1998)]{tripp98} {Tripp, T. M., Lu, L., \&
Savage, B. D. 1998} \textit{ApJ}, 508, 200

\bibitem[Tripp \& Savage (2000)]{tripp00a} {Tripp, T. M., \& Savage,
B. D.} 2000, \textit{ApJ}, 542, 42

\bibitem[Tripp, Savage, \& Jenkins (2000)]{tripp00} {Tripp, T. M.,
Savage, B. D., \& Jenkins, E. B.} 2000, \textit{ApJ}, 534, L1


\bibitem[Tripp et al. (2002)]{tripp02} {Tripp, T. M., et al.} 2002,
\textit{ApJ}, 575, 697

\bibitem[Tumlinson et al. (2005)]{tum05} {Tumlinson, J., Shull, J. M.,
Giroux, M. L., \& Stocke, J. T.} 2005, \textit{ApJ}, 620, 95

\bibitem[Tytler et al. (1995)]{tyt95} {Tytler, D., Fan, X. M., Burles,
S., Cottrell, L., Davis, C., Kirkman, D. \& Zuo, L.} 1995, in
\textit{QSO Absorption Lines}, ed. G. Meylan (Berlin: Springer), p289

\bibitem[Williger et al. (2005)]{willi05} {Williger, G. M., Heap,
S. R., Weymann, R. J., Dav\'{e}, R., Ellison, E., Carswell, R. F.,
Tripp, T. M., \& Jenkins, E. B.} 2005, \textit{ApJ}, in press
(astro-ph/0505586)


\end{thebibliography}
\end{document}